\documentclass[twocolumn,secnumarabic,amssymb, nobibnotes, aps, prl]{revtex4-1}
\usepackage{amssymb, graphicx, latexsym, amsfonts, bm}

\begin{document}

\title{Non-modal Simon-Hoh instability of a plasma with a shearing Hall current}

\author{V. V. Mikhailenko}\email[E-mail: ]{vladimir@pusan.ac.kr}
\affiliation{Pusan National University,  Busan 609--735, S. 
Korea.}

\author{V. S. Mikhailenko}\email[E-mail: ]{vsmikhailenko@pusan.ac.kr}
\affiliation{Pusan National University,  Busan 609--735, S. Korea.}

\author{Hae June Lee}\email[E-mail: ]{haejune@pusan.ac.kr}
\affiliation{Pusan National University,  Busan 
609--735, S. Korea.}

\date{\today}

\begin{abstract}
A new analytical nonmodal approach to investigate the plasma instabilities driven by the sheared current is presented and applied to the analysis 
of the linear evolution of the Simon-Hoh (S-H) instability of a plasma in the inhomogeneous electric field. We found analytically strong nonmodal growth 
for this instability which is missed completely in the normal modes analysis but dominates the normal mode growth when the shearing rate of the current velocity is above the 
growth rate of the S-H instability with uniform current. It includes also the nonmodal growth for the subcritical perturbations, which are suppressed in 
plasma with uniform Hall current. 
\end{abstract}


\maketitle


\textit{Introduction}. -- It is acknowledged last time that the usually applied normal mode analysis fails to predict the behaviour of the instabilities in 
sheared flows\cite{Trefethen, Schmid, Farrell, Mikhailenko-2000, Schmid2, Camporeale, Friedman, Squire}, particularly in predicting the temporal evolution 
and stability of the shearing flows in short-time limit and the onset of subcritical turbulence in sheared flows. The origin of this problem is in the 
assumption that the perturbations imposed on an equilibrium flow have a static structure of a plane wave $\sim exp\left(i\mathbf{kr}-i\omega t\right)$ not 
changing with time. The modal approach neglects by the obvious physical effect when it applied to the shearing flow: the perturbations are convected 
by the shearing flow and experience the  distortion by the shearing flow. This distortion is growing with time and forms the 
time dependent process, which is governed by the flow shearing. This effect is accounted for by the non-modal theory, grounded on the methodology of the 
shearing modes \cite{Mikhailenko-2000, Squire}. These modes are the waves that have a static spatial structure in the frame convected with sheared fluid or 
plasma flow. They are observed in the laboratory frame as the perturbations with time dependent structure determined by the flow shearing
\cite{Mikhailenko-2016}. The temporal evolution of the shearing modes is investigated by the employment of the initial value scheme, which does not impose 
\textit{a priori} any constraints on the form that solution may take\cite{Mikhailenko-2000, Squire, Mikhailenko-2016, Mikhailenko-2018}.  

The principal difference between fluids and plasmas is the possibility of the relative motion  of plasma species (electrons and ions), i.e. of the 
current in the applied electric and magnetic fields. The current driven instabilities of a plasma are the most ubiquitous in space, controlled 
fusion, and laboratory plasmas. Generally, the electric and magnetic fields are inhomogeneous in plasmas and the current velocity formed by the relative 
motion of plasma species is spatially inhomogeneous and sheared. The investigations of the current velocity shearing effect on the temporal evolution of the 
instabilities, turbulence and anomalous transport in plasmas with inhomogeneous electric field are completely missing to date. 

The current across the magnetic field formed by the relative motion of the magnetized electrons and unmagnetized ions in crossed electric and magnetic fields  
(generally referred as Hall current) is the basic feature of the Hall plasmas, widely employed in numerous $E\times B$ devices. The theory of the 
instabilities driven by the Hall current and resulted turbulence and anomalous transport of the Hall plasmas grounds on the application of the canonical 
normal modes approach. The investigation of the effect of the  Hall current shearing, formed by the inhomogeneity of the electric and magnetic fields, on the 
temporal evolution, transient growth and possibility of the development of 
the subcritical turbulence can't be performed with employment of the modal approach and is absent yet. In this letter, we present analytical 
nonmodal approach grounded on the methodology of the shearing modes to the investigating  the temporal evolution of the instabilities driven by the sheared 
current. We consider as a sample the hydrodynamic gradient-drift electrostatic Simon-Hoh (S-H) instability \cite{Simon, Hoh} of the collisionless Hall plasma 
\cite{Sakawa} in the inhomogeneous electric field. The developed theory treats the linear stability as an initial value problem and allows to discover and 
analytically investigate the transient non-modal phenomena in plasma with a sheared current which are missed completely in the normal modes approach. We 
found  that the current shearing is the source of the fast strongly nonmodal growth of the perturbations, which dominates the linear modal growth in the 
plasma with shearless current, when the velocity shearing rate is above the growth rate of 
the S-H instability of plasma with uniform Hall current. It provides also the non-modal growth for the subcritical perturbations, which are suppressed in 
plasma with uniform Hall current. 

\textit{The nonmodal theory of the S-H instability driven by the shearing current}. -- We analyse the stability of a plasma immersed to the nonuniform 
electric field $\mathbf{E}_{0}=E_{0}\left(x\right)\mathbf{e}_{x}$  directed across the uniform magnetic field $\mathbf{B}=B\mathbf{e}_{z}$ pointed along 
the coordinate $z$ as an initial value problem without application any spectral transforms in time variable. The model of the Hall plasmas with
unmagnetized ions and magnetized electrons is applicable to the processes the temporal evolution of which is limited by the period of the ion cyclotron 
Larmor motion $\omega_{ci}^{-1}$, and the strength of the magnetic field is such 
that the ion Larmor radius $\rho_{i}$ is much larger than the characteristic plasma length scale $L$, whereas the electron Larmor radius $\rho_{e}$ much 
less than $L$. The dynamics  of the electron component is determined by 
the electron continuity equation for the perturbation $n_{1e}\left(x,y,t 
\right)$ of the inhomogeneous along the coordinate $x$ equilibrium electron density $n_{0e}\left(x\right)$,
\begin{eqnarray}
&\displaystyle 
\frac{\partial n_{1e}}{\partial t}+V_{0e}\left( x\right)\frac{\partial n_{1e}}{\partial y} 
=\frac{c}{B_{0}}\frac{\partial\phi }{\partial y}\frac{\partial n_{0e}}{\partial x},
\label{1}
\end{eqnarray}
in which the nonuniform equilibrium velocity  $\mathbf{V}_{0e}\left( x\right)$ and the perturbed velocity $\mathbf{v}_{1e}\left(x,y,t\right)$ are 
determined  as the velocities of $\mathbf{E}\times\mathbf{B}$ drift, i.e.
\begin{eqnarray}
&\displaystyle 
\mathbf{V}_{0e}\left( x\right) =-\frac{c}{B_{0}}E_{0}\left(x\right)\mathbf{e}_{y},
\label{2}
\\ 
& \displaystyle
\mathbf{v}_{1e}\left(x,y,t\right)=-\frac{c}{B_{0}}\frac{\partial\phi }{\partial y} \mathbf{e}_{x}+\frac{c}{B_{0}}\frac{\partial\phi }{\partial x} 
\mathbf{e}_{y},
\label{3}
\end{eqnarray}
where $\phi$ is the potential of the electrostatic perturbations. Equation (\ref{1}) contains the spatial inhomogeneity, introduced by the sheared electron 
flow velocity $V_{0e}\left( x\right)$. This inhomogeneity is excluded from Eq. (\ref{1}) by the transformation of Eq. (\ref{1}) to the coordinates $\xi, 
\eta$ convected with sheared electron flow,
\begin{eqnarray}
&\displaystyle 
\xi=x, \quad \hat{t}=t, \quad \eta =y-V_{0e}\left(x\right)t.
\label{4}
\end{eqnarray}
With these coordinates the electron continuity equation (\ref{1}) has simple solution
\begin{eqnarray}
&\displaystyle 
n_{1e}\left(\xi,\eta, \hat{t} \right) =n_{1e}\left(\xi,\eta,t_{0} \right)
\nonumber
\\ 
& \displaystyle
+\frac{c}{B_{0}}\frac{dn_{0e}}{d\xi}
\int\limits^{\hat{t}}_{t_{0}}d\hat{t}_{1}\frac{\partial }{\partial \eta}\psi\left(\xi,\eta,\hat{t}_{1} \right),
\label{5}
\end{eqnarray}
where $\psi\left(\xi,\eta, \hat{t}_{1} \right)=\phi\left(x, y-V_{0e}\left(x\right)t_{1}, t_{1} \right)$. 

The governing equations for the cold unmagnetized ions are the linearised ion momentum equation for the perturbed 
ion velocity $\mathbf{v}_{i}$, with solution 
\begin{eqnarray}
&\displaystyle 
\mathbf{v}_{i}=- \frac{e}{m_{i}}\int\limits^{t}_{t_{0}}\nabla \phi\left(x,y,t_{1} \right) dt_{1},
\label{6}
\end{eqnarray}
and the linearised ion continuity equation
\begin{eqnarray}
&\displaystyle 
\frac{\partial n_{i}}{\partial 	t} + n_{0i}\nabla\cdot\mathbf{v}_{1i}=0,
\label{7}
\end{eqnarray}
for the perturbation $n_{1i}\left(x,y,t \right)$ with solution
\begin{eqnarray}
&\displaystyle 
n_{1i}\left(x,y,t \right)-n_{1i}\left(x,y,t_{0} \right)
\nonumber
\\ 
& \displaystyle
=\frac{en_{0i}}{m_{i}}
\int\limits^{t}_{t_{0}}dt_{1}
\int\limits^{t_{1}}_{t_{0}}dt_{2}\nabla^{2} \phi\left(x,y,t_{2} \right). 
\label{8}
\end{eqnarray}
The contributions of the equilibrium velocity $\mathbf{V}_{0i}$ and of the inhomogeneity of the ion density $n_{0i}\left(x\right)$ to $n_{1i}$ are in $
\left(\omega_{ci}t\right)^{-1}\gg 1$ times less than the corresponding contributions from electrons to $n_{1e}$ and are neglected in Eq. (\ref{8}) in 
model of plasma with unmagnetized ions. Employing the approximation of the quasi-neutrality  $n_{1i}\left(x,y,t \right)=n_{1e}\left(\xi,\eta, \hat{t}\right)$, 
which is valid here for the evolution time limited by the interval $\omega_{pi}^{-1}< t-t_{0}<\omega_{ci}^{-1}$ 
which exists for the Hall plasma devices, where the ion plasma frequency $\omega_{pi}$ is usually much above the ion cyclotron frequency $\omega_{ci}$,
we obtain the  governing integral equation for the electrostatic potential $\phi\left(x,y,t \right)$ Fourier transformed over coordinates $x, y$ ,
\begin{eqnarray}
&\displaystyle 
k^{2}_{\bot}v^{2}_{s}\int\limits^{t}_{t_{0}}dt_{1}\left(t-t_{1}\right)\phi\left( k_{x}, k_{y}, t_{1}\right)=ik_{y}v_{de}\int\limits^{t}_{t_{0}}dt_{1}
\nonumber
\\ 
& \displaystyle
\times\int\limits_{-\infty}^{\infty}dx
e^{-ik_{x}x-ik_{y}V_{0}\left(x \right)\left( t-t_{1}\right) }\phi
\left(x, k_{y}, t_{1} \right),
\label{9}
\end{eqnarray}
where $k^{2}_{\bot}=k^{2}_{x}+k^{2}_{y}$, $v_{s}=\left(T_{e}/m_{i}\right)^{1/2}$ is the ion sound velocity, $v_{de}=\left(-cT_{e}/|e|B\right)
d\ln n_{e0}/dx$ is the electron diamagnetic velocity and the relation 
\begin{eqnarray}
&\displaystyle 
\psi\left(\xi, k_{y}, \hat{t}_{1}\right) =e^{ik_{y}V_{0}\left(x \right) t_{1}}\phi\left( x, k_{y}, t_{1}\right),
\label{10}
\end{eqnarray}
which determines the Fourier transform of the potential $\psi\left(\xi, \eta, \hat{t}_{1}\right)$ over variable $y$, with Fourier transform $\phi\left( x, 
k_{y}, t_{1} \right)$ of the potential $\phi\left( x, y, t_{1} \right)$, was used. The simplest solution to this equation is obtained with canonically 
applied local approximation which assumes that all modes being considered have wavelength much smaller than the spatial scale length $L_{v}$ of the flow 
velocity inhomogeneity, i.e. $L_{v}k_{x}\gg 1$. With this approximation the solution to Eq. (\ref{9}) is of the modal 
form, $\phi\left( x, y, t \right) =\hat{\phi}\left( x, y\right)e^{-i\omega t}$ , where $\omega$ is determined by the dispersion equation \cite{Romadanov} 
\begin{eqnarray}
&\displaystyle 
\frac{k^{2}_{\bot}}{\omega^{2}}=\frac{k_{y}v_{de}}{\omega-k_{y}V_{0}\left(x\right)},
\label{11}
\end{eqnarray}
obtained by the Fourier transform in time of Eq. (\ref{9}) with $t_{0}\rightarrow -\infty$ limit explored by the eigenmode analysis. The solution to this 
equation 
\begin{eqnarray}
&\displaystyle 
\omega \left(\mathbf{k}, x\right)= \frac{k^{2}_{\bot}v^{2}_{s}}{2k_{y}v_{de}}\left(1\pm \left(1-4\frac{V_{0}\left(x \right)}{v_{de}}\frac{k^{2}_{y}v^{2}
_{de}}{k^{2}_{\bot}v^{2}_{s}}\right)^{1/2}\right)
\label{12}
\end{eqnarray}
predicts that S-H instability develops when \\$4V_{0}\left(x \right)/v_{de}>k^{2}_{\bot}v^{2}_{s}/k^{2}_{y}v^{2}_{de}$. 

Consider now the limits of applicability of the local approximation employed for the inhomogeneous electron flow velocity $V_{0}\left(x\right)$ 
in the modal solution (\ref{12}) to Eq. (\ref{9}). With the Taylor expansion of the electron flow velocity, $V_{0}\left(x \right)=V_{0}^{(0)}+V'_{0}x$,  the 
exponent in the rhs of Eq. (\ref{9}) becomes equal to $e^{-i\left(k_{x}+V'_{0}k_{y}\left(t-t_{1}\right)\right)x-ik_{y}V^{(0)}_{0}\left(t-t_{1}\right)}$. 
It displays that the wave number component $k_{x}$ in the electron term changes onto the time dependent component $\hat{k}_{x}= k_{x}+k_{y}V'_{0}\left(t-
t_{1} \right) $ in the case of the sheared flow. The local approximation is valid when the time dependent term which contains the flow velocity shear may be 
neglected, i.e. when $|k_{x}|\gg |
k_{y}V'_{0}\left(t-t_{1}\right)|$. With the estimate $V'_{0}\sim V_{0}/L_{v}$ it holds for the limited time, when $k_{x}L_{v}\gg |k_{y}V_{0}\left(t-t_{1}
\right)|$ and may be violated at the longer time for which $|V'_{0}k_{y}\left(t-t_{1}\right)|>k_{x}$. Therefore, the condition $k_{x}L_{v}\gg 1$  of the 
"slow variation of $V_{0}\left( x\right)$ on the 
wavelength scale" is insufficient for the application of the local approximation for the inhomogeneous flow velocity $V_{0}\left(x\right)$ and employing 
the modal approach to the investigations of the stability of a plasma with sheared current velocity. For the time $t-t_{0}$ of the order of the inverse 
growth rate $\gamma^{-1}$ of the linear modal instability, the non-modal effects develop before the modal linear instability and  determine the temporal 
evolution of the plasma perturbations when 
\begin{eqnarray}
&\displaystyle 
|V'_{0}|> \frac{k_{x}}{k_{y}}\gamma.
\label{13}
\end{eqnarray}
For the considered here the S-H instability this condition is satisfied when 
\begin{eqnarray}
&\displaystyle 
|V'_{0}|> \frac{k_{x}}{k_{y}}k_{\bot}v_{s}\left(\frac{V_{0}^{(0)}}{v_{de}}\right)^{1/2}.
\label{14}
\end{eqnarray}
This condition demands that nonmodal methods should be devised and applied to examine the responses of such sheared currents to plasma perturbations. 

Equation (\ref{14}) predicts also that the nonmodal effects are negligible and the local approximation for the inhomogeneous velocity $V_{0}\left(x\right) 
$ is admissible only for the sufficiently large $V_{0}^{(0)}$,
\begin{eqnarray}
&\displaystyle 
V_{0}^{(0)}> v_{de} \frac{k^{2}_{y}}{k^{2}_{x}}\left(\frac{V'_{0}}{k_{\bot}v_{s}}\right)^{2}
\label{15}
\end{eqnarray}
for which the growth rate of the S-H instability is larger than the velocity shearing rate.

Equation (\ref{9}) Fourier transformed over $x$ with accounted for the shear $V'_{0}$ of the electron flow velocity  becomes
\begin{eqnarray}
&\displaystyle 
k^{2}_{\bot}v^{2}_{s}\int\limits^{t}_{t_{0}}dt_{1}\left(t-t_{1}\right)\phi\left( k_{x}, k_{y}, t_{1}\right)
\nonumber
\\ 
& \displaystyle
=ik_{y}v_{de}\int\limits^{t}_{t_{0}}dt_{1}
e^{-ik_{y}V^{(0)}_{0}\left( t-t_{1}\right) }
\nonumber
\\ 
& \displaystyle
\times\phi\left(k_{x}+k_{y}V'_{0}\left( t-t_{1}\right), k_{y}, t_{1} \right).
\label{16}
\end{eqnarray}
Due to the electron sheared motion, the separate spatial Fourier harmonic $\phi\left( k_{x},k_{y},t_{1}\right)$ of the potential in the ion 
frame of references is observed in the sheared electron flow as a {\textit{Doppler-shifted sheared mode}} $e^{-i\hat{k}_{y}V_{0}^{\left(0 \right) }\left(t-
t_{1}\right)}\phi\left(\hat{k}_{x} , \hat{k}_{y}, t_{1} \right)$ with wave number $\hat{k}_{y}=k_{y}$ and time-dependent wave number $
\hat{k}_{x}=k_{x}+k_{y}V'_{0}\left(t-t_{1}\right)$. In the normal modes approach, the effect of the mode shearing is  missed.  The modal S-H instability  
(\ref{12}) develops when the Doppler-shifted frequency $k_{y}v_{de}-k_{y}V_{0}$ of the electron beam mode becomes negative. We found, however, that the 
current velocity shear itself is a source of the current driven instabilities of new nonmodal type. Due to continuous distortion of the wave pattern observed in frame 
convected with electron 
component, the interaction of electrons with electrostatic ion density perturbation becomes time-dependent. This effect results 
in the non-modal time dependence of the perturbed potential, which is determined by the solution of integral equation (\ref{16}). 
Here we present simple analytical solution to this equation, which displays strongly nonmodal behaviour of the S-H instability at time interval 
$\omega_{0}^{-1}\ll t-t_{0} < \left(V'_{0}\right)^{-1}$ assuming that $V'_{0}$ is much less than the frequency $\omega_{0}=k^{2}_{\bot}v^{2}_{s}/k_{y}v_{de}$ of the S-H instability.  This assumption is compatible with condition (\ref{14}) for $k_{\bot}v_{s}>k_{x}\left(V_{0}v_{de}\right)^{1/2}$. We 
will find the approximate solution to this equation in the Wentzel-Kramers-Brillouin (WKB)-like form,
\begin{eqnarray}
&\displaystyle 
\phi\left(k_{x},k_{y},t_{1} \right) =\Phi\left(k_{x},k_{y} \right)
e^{-i \int\limits^{t_{1}}_{t_{0}}\omega\left(k_{x}, k_{y}, t_{2} \right)dt_{2}}, 
\label{17}
\end{eqnarray} 
where $\Phi\left(k_{x},k_{y} \right)=\int \limits ^{\infty}_{-\infty}e^{-ik_{x}x-ik_{y}y}\phi\left(x, y, t_{0} \right)dxdy$ is the Fourier 
transform of the initial perturbation of the potential at $t=t_{0}$. The solution (\ref{17}) is observed in the electron frame of 
references as a shearing mode
\begin{eqnarray}
&\displaystyle 
\phi\left(\hat{k}_{x},\hat{k}_{y},t_{1} \right) 
=\Phi\left(\hat{k}_{x},\hat{k}_{y} \right)
\nonumber
\\ 
& \displaystyle
\times e^{-i \int\limits^{t_{1}}_{t_{0}}\left(\omega\left(\hat{k}_{x}, \hat{k}_{y}, t_{2} \right)-\hat{k}_{y}V_{0}^{(0)}\right)dt_{2}}. 
\label{18}
\end{eqnarray}
The frequency $\omega\left(k_{x}, k_{y}, t_{1} \right)$ changes with time due to the  
electron flow velocity shear $V'_{0}$ and is identical to frequency (\ref{12}) when $V'_{0}=0$. The solution for $\omega\left(k_{x}, k_{y}, t_{1} \right)$ 
will be derived in the form of the power series expansion in powers of $V'_{0}/\omega_{0}<1$.  By the integration by parts of Eq. (\ref{16}) we obtain the expansion, 
\begin{widetext}
\begin{eqnarray}
&\displaystyle 
e^{-i\int\limits^{t}_{t_{0}}dt_{1}\omega\left(k_{x},k_{y},t_{1} \right)}\left\{-\frac{k^{2}_{\bot}v^{2}_{s}}{\omega_{0}^{2}\left(k_{x},k_{y}\right)}
+\frac{k_{y}v_{de}}{ \omega_{0}\left(k_{x},k_{y}\right)-k_{y}V_{0}^{\left(0\right) }}
-k^{2}_{\bot}v^{2}_{s}\frac{3i}{\omega^{4}_{0}\left(k_{x},k_{y}\right)} \frac{d\omega}{dt}
\right.
\nonumber
\\ 
& \displaystyle
\left.+\frac{ik_{y}v_{de}k_{y}V'_{0}}
{\left(\omega_{0}\left(k_{x},k_{y}\right)-k_{y}V_{0}^{(0)}\right)^{2} }\left(\frac{\partial 
\omega}{\partial k_{x}}\left(t-t_{0}\right)+\frac{\partial }{\partial k_{x}}\ln\Phi\left(k_{x},k_{y}\right)+\frac{\partial 
\omega}{\partial k_{x}}\left(t-t_{0}\right)\right)\right.\nonumber
\\ 
& \displaystyle
\left.+\frac{ik_{y}v_{de}}{\left(\omega_{0}\left(k_{x},k_{y}\right)-k_{y}V_{0}^{(0)}\right)^{3}}\left[\frac{\partial}{\partial t}\omega\left(k_{x},k_{y},t
\right)+k_{y}V'_{0}\frac{\partial }{\partial k_{x}}\omega_{0}\left(k_{x},k_{y}\right)\right] \right\}
\nonumber
\\ 
& \displaystyle -ik_{y}v_{de}A\left(k_{x}, k_{y}, t-t_{0}, t_{0}\right)\Phi\left(k_{x}+k_{y}V'_{0}\left(t-t_{0}\right), k_{y}\right)=0,
\label{19}
\end{eqnarray} 
\end{widetext}
in which the terms of the order of $O\left( \left(V'_{0}/\omega_{0}\right)^{2}\right)$ are neglected. In Eq. (\ref{19})
\begin{eqnarray}
& \displaystyle
A\left(k_{x}, k_{y}, t-t_{0}, t_{0}\right)=
\nonumber
\\ 
& \displaystyle
-\frac{i}{\omega\left(k_{x}+k_{y}V'_{0}\left(t-t_{0}\right),k_{y},t_{0} \right)-k_{y}V_{0}^{(0)}}
\nonumber
\\ 
& \displaystyle
+\frac{k_{y}V'_{0}\frac{\partial }{\partial k_{x}}\ln\Phi\left(k_{x}+k_{y}V'_{0}\left(t-t_{0}\right),k_{y}\right)}{\left(\omega\left(k_{x}+k_{y}V'_{0}
\left(t-t_{0}\right),k_{y},t_{0} \right)-k_{y}V_{0}^{(0)}\right)^{2}}
\label{20}
\end{eqnarray} 
contains the input from the $t=t_{0}$ limit of the integration of Eq. (\ref{16}) by parts. This term is much less than the first term for the exponentially 
growing perturbations and may be neglected in such a case. The time dependence of $\omega\left(k_{x},k_{y},t \right)$ originates from the electron sheared 
motion and is slow  when $V'_{0}\omega_{0}\ll \omega_{0} $, i. e. 
\begin{eqnarray}
& \displaystyle
\frac{\partial\omega}{\partial t}\sim V'_{0}\omega_{0}\ll \omega_{0}^{2}.
\label{21}
\end{eqnarray}
The equation for $\omega\left(k_{x},k_{y},t \right)$ follows from Eq. (\ref{19}),
\begin{eqnarray}
&\displaystyle 
\frac{\partial\omega}{\partial t}=ik_{y}V'_{0}\,\frac{\partial \omega^{2}_{0}\left(k_{x}, k_{y}\right)}{\partial k_{x}}
\nonumber
\\ 
& \displaystyle
-2k_{y}V'_{0} \omega_{0}\frac{\partial }{\partial k_{x}}\ln\Phi\left(k_{x}, k_{y} \right)-4k_{y}V'_{0}\,\frac{\partial \omega_{0}}{\partial 
k_{x}},
\label{22}
\end{eqnarray} 
where the relation $\omega_{0}-k_{y}V_{0}^{(0)}=k_{y}v_{de}\omega_{0}^{2}/k^{2}_{\bot}v^{2}_{s}$ for $\omega_{0}$ was used. This equation is solved with initial value $
\omega\left(k_{x},k_{y},t=t_{0} \right)=\omega_{0}\left(k_{x},k_{y} \right)$, where $\omega_{0}
\left(k_{x},k_{y} \right)$ is determined by  Eq. (\ref{12}). The solution  to Eq. (\ref{22}) displays, that the  S-H instability in the sheared 
electron flow is a non-modal under condition (\ref{14}) and is rapidly growing for $k_{y}V'_{0}>0$ with time dependent growth rate and frequency,
\begin{widetext}
\begin{eqnarray}
&\displaystyle 
-i\int\limits^{t}_{t_{0}}\omega\left(k_{x}, k_{y}, t_{1} \right)dt_{1} =-i\omega_{0}\left(k_{x}, k_{y}\right)\left(t-t_{0}\right)+k_{y}V'_{0}\left[\frac{1}
{2}\frac{\partial}{\partial k_{x}}\left(\omega^{2}_{0}
\left(k_{x}, k_{y}\right)\right)\left(t-t_{0}\right)^{3}\right.
\nonumber
\\ 
& \displaystyle
\left. +i\omega_{0}\left(k_{x}, k_{y}\right)\left(t-t_{0}\right)^{2}\frac{\partial }{\partial k_{x}}\ln\Phi\left(k_{x}, k_{y} \right)+2i\frac{\partial 
\omega_{0}\left(k_{x},k_{y} \right)}{\partial k_{x}}\left(t-t_{0}\right)^{2}\right]. 
\label{23}
\end{eqnarray} 
\end{widetext}
Eq. (\ref{23}) displays that the rapid nonmodal growth of the electrostatic potential,
$|\phi \left(k_{x},k_{y}, t \right)|=\phi \left(k_{x},k_{y}, t_{0} \right)\exp\left(k_{y}V'_{0}\frac{1}
{2}\frac{\partial}{\partial k_{x}}\left(\omega^{2}_{0}\left(k_{x}, k_{y}\right)\right)\left(t-t_{0}\right)^{3}\right)$, occurs  for the perturbations which 
propagate in the direction for which $k_{y}V'_{0}>0$, and the strong nonmodal damping occurs for the perturbations for which $k_{y}V'_{0}<0$.  It is 
important to note that nonmodal solution (\ref{23}) predicts the growth of  the subcritical perturbations, which damp out in plasma with uniform Hall 
current, but for which $k_{y}V'_{0}>0$ and condition (\ref{14}) holds. It may be anticipated that the plasma turbulence resulted from the development of the 
nonmodal S-H instability will be nonmodal with unknown yet unusual properties.

\textit {Conclusions} -- We found that the inclusion of the nonmodal analysis is necessary when condition (\ref{14}) holds. This analysis completely changes 
the predictions of the modal linear theory of the S-H instability which is generally employed for the plasmas in inhomogeneous electric field. We 
demonstrate analytically that when the current velocity shearing rate is above the growth rate of the modal S-H instability driven by the uniform current the 
strong rapidly growing non-modal instability develops which dominates the development of the normal mode instability.  This instability includes also the 
nonmodal growth for the subcritical perturbations, which are suppressed in plasma with uniform Hall current. It is important to note, that  condition 
(\ref{13})  is a general condition for the inapplicability of the local approximation for the inhomogeneous current velocity and of the dominance of the 
nonmodal effects over the modal ones for the current driven instabilities with sheared current velocity. 

This work was supported by National R$\&$D Program through the National Research Foundation of Korea (NRF) funded by the Ministry of Education, Science 
and Technology (Grant No. NRF-2017R1A2B2011106) and BK21PLUS Creative Human Resource Development Program for IT Convergence.


\begin{thebibliography}{}
\bibitem{Trefethen} L. N. Trefethen, A. E. Trefethen, S. C. Reddy, and T. A. Driscoll, Science {\bf 261}, 578 (1993).
\bibitem{Schmid} P. J. Schmid, Annu. Rev. Fluid Mech. {\bf 39}, 129 (2007).
\bibitem{Farrell} B. F. Farrell , J. Atmos. Sci. {\bf 39}, 1663 (1982).
\bibitem{Mikhailenko-2000} V. S. Mikhailenko, V. V. Mikhailenko, and K. N. Stepanov, Phys. Plasmas {\bf 7}, 94 (2000).
\bibitem{Schmid2} P. J. Schmid,  Phys. Plasmas {\bf 7}, 1788 (2000).
\bibitem{Camporeale} E. Camporeale, T. Passot, D. Burgess, Astrophys. J. {\bf 715}, 260 (2010).
\bibitem{Friedman} B. Friedman, T. A. Carter, Phys.Rev.Lett. {\bf 113}, 025003 (2014).
\bibitem{Squire} J. Squire, A. Bhattacharjee, Phys. Rev. Lett. {\bf 113}, 025006 (2014).
\bibitem{Mikhailenko-2016} V. V. Mikhailenko, V. S. Mikhailenko, Hae June Lee, Phys. Plasmas {\bf 23}, 062115 (2016).
\bibitem{Mikhailenko-2018} V. S. Mikhailenko, V. V. Mikhailenko, Hae June Lee, Phys. Plasmas {\bf 25}, 012902 (2018).
\bibitem{Choueiri} E. Y. Choueiri, Phys. Plasmas {\bf 8}, 1411 (2001).
\bibitem{Simon} A. Simon, Phys Fluids  {\bf 6}, 382 (1963). 
\bibitem{Hoh} F. C. Hoh, Phys Fluids  {\bf 6}, 1184 (1963). 
\bibitem{Sakawa} Y. Sakawa, C. Joshi, P. K. Kaw, F. F. Chen, V. K. Jain, Phys.Fluids B{\bf 5}, 1681 (1993).
\bibitem{Romadanov} I. Romadanov, A. Smolyakov, Y. Raitses, I. Kaganovich, T. Tian, S. Ryzhkov, Phys. Plasmas {\bf 23}, 122111 (2016).
\end{thebibliography}
\end{document}